\documentclass[11pt,draftcls,onecolumn]{IEEEtran}
\usepackage{etex}
\usepackage{amsmath,graphicx,amssymb,epsfig,setspace,multicol,cite,float,dsfont}
\usepackage[latin1]{inputenc}
\usepackage{graphicx}

\usepackage{schemabloc}  
\usepackage{epsfig}       
\usepackage{pst-all}      

\usepackage{tikz}
\usepackage{epstopdf}
\usetikzlibrary{positioning}

\DeclareTextSymbol{\degre}{T1}{6}
\DeclareTextSymbol{\degre}{OT1}{23}

\newcommand\diag{{\rm Diag}}
\newcommand\e{{\rm E}}
\newcommand\pardef{ \stackrel{{\rm def}}{=} }

\def\QED{\mbox{\rule[0pt]{1.4ex}{1.4ex}}}

\usepackage{color}
\usepackage{soul}

\begin{document}

\title{
On linear MMSE receivers in the presence of data-like multiuser interference}

\author{Pascal Chevalier, Jean-Pierre Delmas, and Roger Lamberti

\thanks{
P. Chevalier is with CNAM, laboratoire CEDRIC, HESAM University,
292 rue Saint Martin, 75141 Paris Cedex 3, France
and also with
Thales  France, HTE/AMS/TCP, 4 Av. Louvresses, 92622 Gennevilliers, Cedex, France
e-mail: pascal.chevalier@cnam.fr, pascal.chevalier@thalesgroup.com.

J.-P. Delmas is with Samovar laboratory, Telecom SudParis, Institut Polytechnique de Paris,
91120 Palaiseau, France,
e-mail: jean-pierre.delmas@it-sudparis.eu.

R. Lamberti is with CITI department, Telecom SudParis, Institut Polytechnique de Paris,
91120 Palaiseau, France,
e-mail: roger.lamberti@orange.fr.
}
}
\maketitle

\begin{abstract}
Linear minimum mean-square error (LMMSE) receivers have been of great interest for many decades for multi-user interference (MUI) mitigation in radiocommunications networks in particular.
Despite the extensive literature on this subject, this topic remains relevant to several current and emerging applications.
In this context, the paper presents a continuous-time approach showing that most of LMME receivers of the literature are sub-optimal. 
In particular, they are typically implemented at the symbol rate following matched filtering to the pulse-shaping filter and require a prior knowledge of the MUI channels, which can be difficult and costly to acquire in practice.
The main contribution of this paper is to address the key challenge of designing new LMMSE receivers that achieve the performance of the optimal receiver without requiring prior knowledge of the MUI channels.
 To this end, two novel receiver LMMSE structures are proposed and analyzed for linearly modulated signals impaired by data-like MUI: a two-input receiver and a three-input receiver. 
 The proposed receivers are shown to be quasi-optimal over a broad range of scenarios. 
 These results provide new perspectives for practical implementation of the optimal LMMSE receiver in the presence of data-like MUI, relying only on knowledge of the signal-of-interest (SOI) channel.
\end{abstract}
\begin{center}
Paper submitted to Signal Processing
\end{center}
%
\section{Introduction}
\label{sec:Introduction}
%
LMMSE receivers have attracted considerable interest for several decades in a wide range of applications, and particularly for mitigating multi-user interference (MUI) in radiocommunication networks. They have been considered, for example, for channel equalization in code-division multiple-access (CDMA) networks \cite{Choi2003,Hadinejad-Mahram2004} and more generally in communication systems \cite{Tuchler2002}, as well as for multi-user detection in CDMA networks \cite{Wang1999,Choi2004,Zhang2005}, in multiple-input multiple-output (MIMO) non-orthogonal multiple-access (NOMA) networks \cite{Lui2019}, and in more general multi-user communication systems \cite{Ibrahim2024}. More recently, LMMSE receivers have continued to attract significant interest in MIMO systems \cite{Park2022}, multi-user multiple-input single-output (MU-MISO) systems \cite{Chen2026}, massive MIMO networks \cite{Tan2022,Chen2023,Zhao2025}, cell-free massive MIMO networks \cite{Bjornson2020,Fu2025}, and decentralized MIMO systems \cite{Zhuang2025}, as well as for orthogonal time-frequency space (OTFS) systems over underwater acoustic channels \cite{Jing2025}, among many other applications.

Despite extensive research on LMMSE receivers over the past several decades, fundamental questions regarding their structural constraints, theoretical performance limits, and practical implementation requirements remain open. 
More specifically, most existing LMMSE receivers operate at the symbol rate following matched filtering with the pulse-shaping filter, thereby imposing a
restrictive structural constraint (sc).
Moreover, conventional designs 
\cite{Choi2003,Hadinejad-Mahram2004,Tuchler2002,Wang1999,Choi2004,Zhang2005,Lui2019,Ibrahim2024,Park2022,Chen2026,Tan2022,Chen2023,Zhao2025,Bjornson2020,Fu2025,Jing2025} fully exploit the common waveform of the signal of interest (SOI) and multi-user interference
(MUI), but require prior knowledge of, or explicit estimation of, both the SOI and MUI channels. Such channel knowledge can be difficult
and costly to acquire in practice. In this context, this paper makes five main contributions.

First, we revisit the conventional sc-LMMSE receiver. We show that,
under the conventional symbol-rate structural constraint, prior
knowledge or estimation of the MUI channels is in fact not required
for its implementation. This receiver is subsequently referred to as
the sc-LMMSE receiver.

Second, we derive the optimal unconstrained LMMSE receiver,
referred to as the o-LMMSE receiver. Unlike the sc-LMMSE receiver,
the o-LMMSE receiver fully exploits the waveform structure and
cyclostationarity of the MUI. Its main limitation, however, is that its
implementation requires knowledge of, or estimation of, the MUI
channels.

Third, we establish the suboptimality of the sc-LMMSE receiver. We
prove that the widely used sc-LMMSE receiver is generally unable to
attain the performance of the o-LMMSE receiver and is therefore
suboptimal in a broad range of practical scenarios. This result
motivates the development of new LMMSE receivers that can
approach the performance of the o-LMMSE receiver without
requiring knowledge of the MUI channels.

Fourth, we propose in this context a novel two-input frequency-shift (FRESH)
LMMSE receiver, referred to as the s-two-input LMMSE receiver. The
proposed receiver avoids the need for MUI channel knowledge by
deliberately modeling the cyclostionnariy MUI as a second-order stationary process.
Despite this simplifying assumption, our analysis shows that the
receiver achieves near-optimal performance over a broad range of
channel and pulse-shaping conditions. Following the same
principle, we also develop a three-input FRESH LMMSE receiver, denoted by s-three-input LMMSE receiver, which provides
a further, albeit modest, performance improvement
over the s-two-input LMMSE receiver.

Fifth, we provide an interpretable analytical performance analysis of
all the considered receivers. To this end, we adopt a continuous-time (CT) framework, 
which preserves conceptual generality, avoids artifacts associated with discrete-time sampling, and enables the
derivation of tractable analytical performance metrics. In particular,
we derive original closed-form expressions for the output SINR of
the considered receivers. 
These expressions provide direct performance comparisons and allow us to establish analytical performance bounds for the SINR given by oversampled discrete-time implementations.

To the best of our knowledge, this approach is novel in that it enables FRESH LMMSE equalizers to achieve near-optimal performance for a broad class of linearly modulated signals without requiring prior knowledge or explicit estimation of the MUI channels.
This result opens new perspectives for the practical implementation of LMMSE receivers in the presence of MUI.
This distinguishes the proposed approach from existing works such as \cite{Chevalier2024}, 
which focuses on widely linear MMSE receivers and is restricted to (quasi-)rectilinear modulations,
and \cite{Gardner1993}, which addresses signal estimation rather
than symbol equalization.

The paper is organized as follows. 
Section \ref{sec:Models and total noise second-order statistics} introduces the observation model and its extensions for two-input and three-input processing, together with the second-order (SO) statistics of the total noise.
Section \ref{sec:M-input LMMSE receivers} derives the sc-LMMSE  and o-LMMSE, as well as the s-$M$-input LMMSE receivers for $M = 1, 2, 3$.
Section \ref{sec:SINR at the output of the LMMSE receivers} gives a common closed-form expression of the SINR on the current symbol at the output of all these derived receivers.
Section \ref{sec:SINR analysis for zero and P MUI} considers scenarios with no MUI and with multiple data-like MUI.
Very insightful interpretable analytical expressions for the corresponding  SINRs are derived for several specific scenarios.
A comprehensive comparative analysis of the derived  SINRs is then performed, both analytically and through computer simulations, for various propagation channels and signal configurations. These results constitute the sixth main contribution of this paper.
Section \ref{sec:BER at the output of the LMMSE receivers for one MUI} shows that the main conclusions drawn from
 the output SINR analysis remain valid when performance is assessed in terms of the output binary error rate (BER). 
 Finally section \ref{sec:Conclusion} concludes this paper.

The following notations are used throughout this paper.
Non boldface symbols denotes scalar, whereas lowercase (uppercase) boldface symbols denote column vectors (matrices). The superscripts $T$, $H$ and $*$ denote the transpose, conjugate transpose and complex conjugate, respectively.  The operator $\otimes$ is the convolution operation.
$\delta(x)$ and $\mathds{1}_{C}(x)$ denote the Kronecker delta and the indicator symbols, respectively, defined as $\delta(x)=1$ for $x=0$
and $\delta(x)=0$ for $x\neq0$, and $\mathds{1}_{C}(x) = 1$ if $x \in C$ and $\mathds{1}_{C}(x) \!=\! 0$ if $x \not\in C$.
The matrices ${\bf 0}_K$ and ${\bf I}_K$ are the zero and the identity matrices of dimension $K$, respectively. 
Moreover, the same notation is used for the Fourier transforms of vectors ${\bf x}$ and
matrices ${\bf X}$, where the time variables $t$ or $\tau$ are replaced by the frequency
variable $f$.
%
\section{Models and total noise second-order statistics}
\label{sec:Models and total noise second-order statistics}
\subsection{Observation model}
\label{sec:Observation model}
We consider an array of $N$ narrowband antennas receiving the contribution of a SOI, $P$ data-like MUI, sharing the same symbol period
and pulse-shaping filter as the SOI, and background noise. The vector of
complex amplitudes at the antenna outputs after frequency synchronization can be written as
\begin{eqnarray}
\nonumber
{\bf x}(t)
&=&
\sum_{\ell}a_{\ell}{\bf g}(t-\ell T)
+
\sum_{1 \le p \le P}\sum_{\ell}a_{p,\ell}{\bf g}_p(t-\ell T)
+{\boldsymbol \epsilon}(t)
\\
\label{eq:model}
&=&
\sum_{\ell}{\bf G}(t-\ell T){\bf a}_{\ell}+{\boldsymbol \epsilon}(t)
\pardef
\sum_{\ell}a_{\ell}{\bf g}(t-\ell T)+{\bf n}(t) \in \mathbb{C}^N.
\end{eqnarray}
The complex symbols $(a_{\ell}, a_{p,\ell})$ are zero-mean independent and identically
distributed random variables drawn from an arbitrary constellation. The vector
${\bf g}(t)= v(t)\otimes{\bf h}(t)$ is the $N\times1$ impulse response of the global SOI
channel, where $v(t)$ and ${\bf h}(t)$ denote the impulse responses of the SOI pulse shaping
filter and propagation channel, respectively.
Similarly, ${\bf g}_p(t) \triangleq v(t)\otimes{\bf h}_p(t)$, where ${\bf h}_p(t)$ denotes the
impulse response vector of the propagation channel associated with the $p$th MUI.
${\bf G}(t)$ is the $N \times (P+1)$ matrix defined by ${\bf G}(t) \pardef [{\bf g}(t), {\bf g}_1(t),...,{\bf g}_p(t)] = v(t) \otimes {\bf H}(t)$
where ${\bf H}(t) \pardef  [{\bf h}(t), {\bf h}_1(t),...,{\bf h}_p(t)]$, ${\bf a}_{\ell}$  is the $(P+1) \times1$ vector defined by
${\bf a}_{\ell} \pardef  [a_{\ell}, a_{1,\ell},..., a_{P,\ell}]^T$.
The vector ${\boldsymbol \epsilon}(t)$ denotes the $N\times1$ background noise,
assumed zero-mean, circular, stationary, and temporally and spatially white. The vector
${\bf n}(t)$ denotes the total noise composed of the $P$ MUIs and the background noise.
Finally, the filter $v(t)$ is assumed to be a normalized (unit-energy) square-root raised
cosine (SRRC) pulse with roll-off factor $\omega$ and bandwidth $B=(1+\omega)/T$.
%
\subsection{Total Noise SO statistics}
\label{sec:Total Noise SO statistics}
The SO statistics of ${\bf n}(t)$ are characterized by the correlation matrix
${\bf R}_n(t, \tau)$ defined by
${\bf R}_n(t, \tau)
\pardef
\e[{\bf n}(t+\tau/2){\bf n}^H(t-\tau/2)]$.
Using \eqref{eq:model}, it is easy to verify that ${\bf R}_n(t, \tau)$ is a periodic functions of $t$ of period $T$. The matrix ${\bf R}_n(t, \tau)$ has then a Fourier series expansion given by
\begin{equation}
\label{eq:defRe}
{\bf R}_n(t, \tau)
=
\sum_{\alpha_{i}}{\bf R}_n^{\alpha_{i}}(\tau)e^{j 2\pi \alpha_{i} t},
\end{equation}
where $\alpha_{i}={i}/T$, $i \in \mathbb{Z}$
are the so-called SO cyclic frequencies of ${\bf n}(t)$.
${\bf R}_n^{\alpha_{i}}(\tau)$
is the  cyclic correlation matrix of ${\bf n}(t)$ for the cyclic frequencies $\alpha_{i}$ and the delay $\tau$, defined by
\begin{equation}
\label{eq:defRea}
{\bf R}_n^{\alpha_{i}}(\tau)
\pardef
<{\bf R}_n(t, \tau)e^{-j 2\pi \alpha_{i} t}>,
\end{equation}
where $<.>$ is the temporal mean operation in $t$ over an infinite observation duration.
The Fourier transform ${\bf R}_n^{\alpha_{i}}(f)$  of
${\bf R}_n^{\alpha_{i}}(\tau)$ is called the  cyclospectrum
of ${\bf n}(t)$ for the cyclic frequencies $\alpha_{i}$.
%
\subsection{Two and three-input extended models for L FRESH processing}
\label{sec:Two and three-input extended models for L FRESH processing}
A conventional linear processing of ${\bf x}(t)$ exploits only the information contained at the zero $(\alpha_0 = 0)$ SO cyclic frequency of ${\bf x}(t)$, through the exploitation of the temporal mean $<{\bf R}_x(t, \tau)>$ of the correlation matrix
${\bf R}_x(t, \tau)
\pardef
\e[{\bf x}(t+\tau/2){\bf x}^H(t-\tau/2)]$
of ${\bf x}(t)$. Consequently, the information contained at the nonzero SO cyclic frequencies of ${\bf x}(t)$ is not exploited.
For small SOI/MUI bandwidths, a large fraction of the signal energy is concentrated at the zero SO cyclic frequency. As the SOI/MUI bandwidth increases, however, the energy contained at the nonzero SO cyclic frequencies $\alpha_i=i/T$, $i\neq0$, increases, while it generally decreases as $|i|$ increases \cite{Gardner1993}, 
\cite{Gardner1987}. Consequently, a conventional LMMSE receiver based solely on the zero cyclic frequency may become highly suboptimal.
To exploit a larger fraction of the residual SO cyclic energy not captured by this receiver, we consider the following two-input FRESH model:
\begin{equation}
\label{eq:model 2 input}
{\bf x}_{2}(t)
\pardef
[{\bf x}^T(t),e^{-j2\pi t/T}{\bf x}^T(t)]^T,
\end{equation}
or equivalently ${\bf x}_{2}(t)
\pardef
[{\bf x}^T(t),e^{j2\pi t/T}{\bf x}^T(t)]^T$.
A two-input processing of ${\bf x}(t)$ exploits the temporal mean $<{\bf R}_{x_2}(t, \tau)>$ of the correlation matrix
${\bf R}_{x_{2}}(t, \tau) \pardef \e[{\bf x}_{2}(t+\tau/2){\bf x}_{2}^H(t-\tau/2)]$ of ${\bf x}_{2}(t)$.
Using decomposition \eqref{eq:defRe}, yields
$<{\bf R}_{x_2}(t, \tau)>=\left( \begin{array}{cc}
\!\!{\bf R}_x^{0}(\tau)\!  &\! e^{-j2\pi\tau/2T}{\bf R}_x^{-1/T}(\tau)\!\!\\
\!\! e^{-j2\pi\tau/2T}{\bf R}_x^{1/T}(\tau)\! &\! e^{-j2\pi\tau/T}{\bf R}_x^{0}(\tau)\!\!\\
\end{array}\right)$ showing that $<{\bf R}_{x_2}(t, \tau)>$ 
captures the information associated with the SO cyclic frequencies $(\alpha_0, \alpha_{-1}, \alpha_1) = (0, -1/T, 1/T)$.
To further exploit the information contained in these frequencies, we consider the following three-input FRESH model:
\begin{equation}
\label{eq:model 3 input}
{\bf x}_{3}(t)
\pardef
[{\bf x}^T(t),e^{-j2\pi t/T}{\bf x}^T(t),e^{j2\pi t/T}{\bf x}^T(t)]^T,
\end{equation}
where 
${\bf R}_{x_{3}}(t, \tau) \pardef \e[{\bf x}_{3}(t+\tau/2){\bf x}_{3}^H(t-\tau/2)]$
now exploits the information contained in $(\alpha_0, \alpha_{-1}, \alpha_1,\alpha_{-2}, \alpha_2) = (0, -1/T, /T,-2/T, 2/T)$.
Increasing the number of inputs further in order to exploit the energy contained at $\alpha_i=i/T$ for $|i|>2$ is generally of limited practical interest, since only a small amount of residual energy is typically present at these higher-order nonzero SO cyclic frequencies. 
In particular for SRRC pulse shaping filters $v(t)$ considered in this paper, the energy of ${\bf x}(t)$ is concentrated in the cyclic frequencies 
$(\alpha_0, \alpha_{-1}, \alpha_1)$.
Finally, it is worth noting that a time-invariant linear processing of ${\bf x}_M(t)$, $(M=2,3)$, corresponds to a time-varying linear processing of the original signal 
${\bf x}(t)$.
%
\subsection{$M$-input generic model for standard and FRESH processing}
\label{sec:M-input generic model for standard and FRESH processing} 
%
With ${\bf x}_1(t)\pardef {\bf x}(t)$ for linear receivers, we deduce from \eqref{eq:model}, \eqref{eq:model 2 input} and \eqref{eq:model 3 input} that 
${\bf x}_M(t)$ always takes the form:
\begin{eqnarray}
\nonumber
{\bf x}_M(t)
&=&
\sum_{\ell}a_{\ell}{\bf g}_M(t-\ell T)
+
\sum_{1 \le p \le P}
\sum_{\ell}\!a_{p,\ell}{\bf g}_{p,M}(t-\ell T)
+{\boldsymbol \epsilon}_M(t)
\\
\label{eq:model M}
&=&
\sum_{\ell}{\bf G}_M(t-\ell T){\bf a}_{\ell}+{\boldsymbol \epsilon}_M(t)
\pardef
\sum_{\ell}a_{\ell}{\bf g}_M(t-\ell T)+{\bf n}_M(t),
\end{eqnarray}
where ${\bf g}_{M}(t)$,
${\bf g}_{p,M}(t)$,
${\boldsymbol \epsilon}_M(t)$
and
${\bf n}_M(t)$
are defined in a similar way as ${\bf x}_{M}(t)$, where
${\bf x}(t)$ is replaced by
${\bf g}(t)$,
${\bf g}_{p}(t)$,
${\boldsymbol \epsilon}(t)$
and
${\bf n}(t)$, respectively, whereas
${\bf G}_1(t)\pardef{\bf G}(t)$ and ${\bf G}_M(t)\pardef [{\bf g}_M(t), {\bf g}_{1,M}(t),...,{\bf g}_{P,M}(t)]$, $M=2,3$.
%
\section{$M$-input LMMSE receivers}
\label{sec:M-input LMMSE receivers}
%
\subsection{MMSE criterion}
\label{sec:MMSE criterion}
%
An $M$-input MMSE receiver considered in this paper generates an output ${\bf y}_M(t)$ which minimizes, at each symbol time $nT$, the MSE criterion defined by 
\begin{equation}
\label{eq:MMSE criterion}
{\rm MSE} 
=\e(|a_n -y_{M}(nT)|^2).
\end{equation}
Note that, in the present work, the MSE is minimized at each symbol time, in contrast to the criterion considered in \cite{Gardner1993}, where the MSE is minimized at every time instant $t$. In other words, the approach adopted in this paper is an equalization approach, whereas that in \cite{Gardner1993} is a signal estimation approach. The different (M)-input receivers considered in this paper differ both in their structures and in the assumptions made regarding the MUI.

\subsection{LMMSE receiver with a particular structure constraint}
\label{sec:LMMSE receiver with a particular structure constraint}
To simplify their implementation, most LMMSE receivers in the literature are implemented at the symbol rate following a matched-filtering operation with the pulse-shaping filter. This implementation imposes a specific structural constraint, which is shown in this paper to be generally suboptimal in a broad range of scenarios. To derive the sc-LMMSE receiver, we denote by
${\bf x}_v(t) \pardef v^*(-t) \otimes {\bf x}(t)$,
the observation vector after matched filtering with the pulse-shaping filter. It then follows from \eqref{eq:model} that, after sampling at the symbol rate, the observation vector is given by
\begin{eqnarray}
\nonumber
{\bf x}_{v}(nT)
&=&
\sum_{\ell}a_{\ell}{\bf g}_{v}((n-\ell) T)
+
\sum_{1 \le p \le P}\sum_{\ell}a_{p,\ell}{\bf g}_{p_{v}}((n-\ell) T)
+{\boldsymbol \epsilon}_{v}(nT)
\\
\label{eq:model v}
&=&
\sum_{\ell}{\bf G}_{v}((n-\ell) T){\bf a}_{\ell}+{\boldsymbol \epsilon}_{v}(nT)
\pardef
\sum_{\ell}a_{\ell}{\bf g}_{v}((n-\ell) T)+{\bf n}_{v}(nT).
\end{eqnarray}
Here,
${\bf g}_{v}(t)$,
${\bf g}_{p_{v}}(t)$,
${\boldsymbol \epsilon}_{v}(t)$
and
${\bf n}_{v}(t)$
are defined in a similar way as ${\bf x}_v(t)$,  
whereas ${\bf G}_{v}(t)$ is defined in a similar way as ${\bf G}(t)$ where
${\bf g}_v(t)$ and ${\bf g}_{p_v}(t)$ replace ${\bf g}(t)$ and ${\bf g}_p(t)$, respectively.
We denote by ${{\bf w}^d}^*(-kT)$, the samples of the discrete time (DT) receiver whose output at time $nT$ is given by 
\begin{equation}
\label{eqfiltrage temps discret}
y(nT)
=
\sum_k
{{\bf w}^d}^{H}(-kT)
{\bf x}_{v}((n-k)T).
\end{equation}
The DT receiver, ${{\bf w}^d}^{*}(-kT)$, whose output \eqref{eqfiltrage temps discret} minimizes \eqref{eq:MMSE criterion} is denoted by 
${{\bf w}_{sc}^d}^{*}(-kT)$ and is called sc-LMMSE receiver. 
Adapting the proof given in \cite{Chevalier2024} for WL MMSE receivers, the frequency response, ${{\bf w}_{sc}^d}^{*}(f)$, of ${{\bf w}_{sc}^d}^{*}(-kT)$ is given by:
\begin{eqnarray}
\nonumber
{\bf w}_{sc}^d(f)
&=&
\pi_a[{\bf R}_{x_{v}}^d(f)]^{-1}{\bf g}_{v}^d(f)
=
\left[1/\pi_a + {{\bf g}_{v}^d}^H(f)[{\bf R}_{n_{v}}^d(f)]^{-1}{\bf g}_{v}^d(f)\right]^{-1}
[{\bf R}_{n_{v}}^d(f)]^{-1}{\bf g}_{v}^d(f)
\\
\label{eq:filter structure}
&\pardef &
{c_{sc}^d}(f)[{\bf R}_{n_{v}}^d(f)]^{-1}{\bf g}_{v}^d(f),
\end{eqnarray}
if $v(f)$ does not vanish in $\Delta \pardef[-1/2T,+1/2T]$. Otherwise ${\bf w}_{sc}^d(f)={\bf 0}$ for the frequencies $f$ which are outside the support of 
${\bf g}_{v}^d(f)\pardef \sum_k {\bf g}_{v}(kT)e^{-j2\pi f k T}$, i.e., such that ${\bf x}_{v}^d(f)\pardef \sum_k {\bf x}_{v}(kT)e^{-j2\pi f k T}={\bf 0}$.
$\pi_a \pardef \e|a^2_k|$ and matrices ${\bf R}_{x_{v}}^d(f)$ and ${\bf R}_{n_{v}}^d(f)$
are the Fourier transforms of matrices
${\bf R}_{x_{v}}^d(kT) \pardef \e[{\bf x}_{v}(nT){\bf x}_{v}^H((n-k)T)]$
and
${\bf R}_{n_{v}}^d(kT) \pardef \e[{\bf n}_{v}(nT){\bf n}_{v}^H((n-k)T)]$,
respectively, where the DT vectors ${\bf x}_{v}(nT)$ and ${\bf n}_{v}(nT)$ are SO stationary.
The output, at time $kT$, of the sc-LMMSE receiver is thus given by   
\begin{eqnarray}
\label{eq:y structure}
y_{sc}(k)
&=&
\int {{\bf w}_{sc}^d}^H(f){\bf x}_{v}(f)e^{j2\pi f kT}df
=
\int c_{sc}^d(f){{\bf g}_{v}^d}^H(f)[{\bf R}_{n_{v}}^d(f)]^{-1}{\bf x}_{v}(f)e^{j2\pi f kT}df.
\end{eqnarray}
Expressions \eqref{eq:filter structure} and \eqref{eq:y structure} show that the sc-LMMSE receiver, subject to the structural constraint commonly adopted in the literature, consists of two successive operations. The first operation, ${{\bf g}_{v}^d}^H(f)[{\bf R}_{n_{v}}^d(f)]^{-1}$, implements a spatio-temporal, or equivalently spatio-frequency, matched filter to the global channel of the SOI in spatially and temporally colored total noise, which is stationary in this case. The second operation, $c_{sc}^d(f)$, implements a scalar discrete-time (DT) filter
corresponding to an LMMSE equalizer for the SOI channel at the output of the matched filter.
Note that the implementation of the receiver in \eqref{eq:filter structure} requires prior knowledge of, or estimation of, ${\bf R}_{x_{v}}^d(f)$ and ${\bf g}_{v}^d(f)$, and hence only of ${\bf h}(kT)$, i.e., the discrete-time impulse response of the SOI channel vector.

\subsection{Optimal LMMSE receiver}
\label{sec:Optimal LMMSE receiver}
To establish the suboptimality of the sc-LMMSE receiver, it is first necessary to derive the optimal LMMSE receiver, hereafter referred to as the o-LMMSE receiver. Unlike the sc-LMMSE receiver, the o-LMMSE receiver is not subject to any structural constraint and fully exploits the waveform structure of the MUI. In particular, it explicitly accounts for the MUI waveform as it appears in the observation model \eqref{eq:model}, thereby inherently exploiting its cyclostationarity. We denote by ${\bf w}^*(-t)$ the continuous-time (CT) receiver whose output is defined by
\begin{equation}
\label{eq:filtre continu}
y(t) 
=
{\bf w}^H(-t) \otimes {\bf x}(t).
\end{equation}
The CT receiver,  ${\bf w}^*(-t)$, whose output \eqref{eq:filtre continu} 
minimizes \eqref{eq:MMSE criterion} at each symbol time, is denoted by ${\bf w}_{o}^*(-t)$.
Adapting the proof given in \cite{Chevalier2024} for WL MMSE receivers, the frequency response ${\bf w}_{o}^*(f)$, of ${\bf w}_{o}^*(-t)$ is such that
\begin{eqnarray}
\nonumber
{\bf w}_o(f)
&=&
{\bf G}(f)[(N_0/\pi_a){\bf I}_{P+1}
+
1/T\sum_{\ell}{\bf G}^H(f-\ell/T){\bf G}(f-\ell/T)]^{-1}{\bf e}_1
\\
\label{eq:filter optimal}
&\pardef&
{\bf G}(f){\bf C}_o^d(f){\bf e}_1
\pardef
{\bf G}(f){\bf c}_o^d(f),
\end{eqnarray}
where $N_0$ is the power spectral density of each component of the noise vector ${\boldsymbol \epsilon}(t)$
and ${\bf e}_1\pardef (1,0,\hdots,0)^T \in \mathbb{R}^{P+1}$.
The output, at time $kT$, of the o-LMMSE receiver is thus given by
\begin{equation}
\label{eq:y optimal}
y_{o}(k)
=\int {\bf w}_{o}^H(f){\bf x}(f)e^{j2\pi f kT}df
=\int {{\bf c}_{o}^d}^H(f){\bf G}^H(f){\bf x}(f)e^{j2\pi f kT}df.
\end{equation}

Expressions \eqref{eq:filter optimal} and \eqref{eq:y optimal} show that the optimal LMMSE receiver consists of three successive operations.
The first operation, ${\bf G}^H(f)$, implements a set of $P+1$ CT matched filters, each matched to the global channel of the SOI or one of the MUI signals.
The second operation consists of sampling the outputs of these $P+1$ matched filters at the symbol rate.
The third operation, ${{\bf c}_{0}^d}^H(f)$, implements a set of $P+1$ DT filters applied to the $P+1$ sampled outputs. This operation jointly performs SOI channel equalization and noise-plus-MUI minimization.
Note that the implementation of the receiver in \eqref{eq:filter optimal}  requires prior knowledge of, or estimation of, $N_0$ and ${\bf G}(t)$, and hence of ${\bf H}(t)$, i.e., the impulse responses of both the SOI and MUI channel vectors. Such channel knowledge may be cumbersome to acquire and may therefore limit the practical implementation of the optimal LMMSE receiver.
%
\subsection{$M$-input LMMSE receivers under the false assumption of SO stationary MUI }
\label{sec:M input LMMSE receivers falsely assuming SO stationary MUI}
%

One approach to designing $M$-input LMMSE receivers $(M=1,2,3)$ that do not require knowledge of, or estimation of, the MUI channels, while imposing no structural constraint, is to estimate only the SOI symbols under the assumption that the MUI signals are second-order stationary. Accordingly, the CT $M$-input receiver 
${\bf w}_{M}^*(-t)$, whose output
$y_M(t) ={\bf w}_M^H(-t) \otimes {\bf x}_M(t)$,
is designed to minimize the MSE criterion in \eqref{eq:MMSE criterion} at each symbol time, assuming that the MUI signals, and consequently the total noise 
${\bf n}_M(t)$ in \eqref{eq:model M} are SO stationary. We denote this receiver by ${\bf w}_{M_s}^*(-t)$ and refer to it as the $M$-input LMMSE receiver under the false stationarity assumption, or, for short, the s-$M$-input LMMSE receiver.
Adapting the proof given in \cite{Chevalier2024} for WL MMSE receivers, the frequency response ${\bf w}_{M_s}^*(f)$ of ${\bf w}_{M_s}^*(-t)$ is given by :
\begin{eqnarray}
\nonumber
{\bf w}_{M_s}(f)
&=&
\Big\{(1/\pi_a)+(1/T)\sum_{\ell}{\bf g}_M^H(f-\ell/T)
[{\bf R}_{n,M}^{0}(f-\ell/T)]^{-1}{\bf g}_M(f-\ell/T)\big\}^{-1}[{\bf R}_{n_M}^{0}(f)]^{-1}{\bf g}_M(f)
\\
\label{eq:filter hypothese stationnaire}
&\pardef &
{c_{M_s}^d}(f)[{\bf R}_{n_M}^{0}(f)]^{-1}{\bf g}_M(f),
\end{eqnarray}
where
${\bf g}_{1}(f)= {\bf g}(f)$,
${\bf g}_{2}(f)\pardef [{\bf g}^T(f),{\bf g}^T(f+\frac{1}{T})]^T$
and
${\bf g}_{3}(f)\pardef [{\bf g}^T(f),{\bf g}^T(f+\frac{1}{T}),{\bf g}^T(f-\frac{1}{T})]^T$.
${\bf R}_{n_M}^0(f)$ which is defined as the Fourier transform of
${\bf R}_{n_M}^0(\tau)\pardef <\e[{\bf n}_M(t+\tau/2){\bf n}_M^H(t-\tau/2)]>$
corresponds to the power spectral density matrix of ${\bf n}_M(t)$.
The output, at time $kT$, of the s $M$-input LMMSE receiver is thus given by  
\begin{equation}
\label{eq:y stationnaire}
y_{M_s}(k)
=
\int {\bf w}_{M_s}^H(f){\bf x}_{M}(f)e^{j2\pi f kT}df
=
\int c_{M_s}^d(f){\bf g}_M^H(f)[{\bf R}_{n_M}^0(f)]^{-1}{\bf x}_{M}(f)e^{j2\pi f kT}df.
\end{equation}
Expressions \eqref{eq:filter hypothese stationnaire}
and \eqref{eq:y stationnaire} show that the LMMSE receivers without any structural constraint, designed under the false assumption that the MUI is second-order stationary, consist of three successive operations.
The first operation, ${\bf g}_M^H(f)[{\bf R}_{n_M}^0(f)]^{-1}$, implements a CT $M$-input spatio-temporal, or equivalently spatio-frequency, pseudo-matched filter to the global channel of the SOI in spatially and temporally colored total noise.
The second operation consists of sampling the output of this pseudo-matched filter at the symbol rate.
The third operation, $c_{M_s}^d(f)$, implements a scalar DT filter corresponding to a pseudo-LMMSE equalizer for the SOI channel at the output of the pseudo-matched filter.
Note that the implementation of the receivers in \eqref{eq:filter hypothese stationnaire} requires prior knowledge of, or estimation of, ${\bf R}_{n}^0(f)$ and ${\bf g}(t)$, and hence of ${\bf h}(t)$, i.e., the impulse response of the SOI channel vector only. Thus, unlike the o-LMMSE receiver, these receivers do not require knowledge or estimation of the MUI channel vectors, which may be advantageous for practical implementation.
%
\subsection{Particular case of an absence of MUI}
\label{sec:Particular case of an absence of MUI}
%
In the absence of MUI, the total noise is SO stationary and thus the s $M$-input LMMSE $(M=1,2,3)$ receivers and the o-LMMSE receivers coincide. 
In this case,
${\bf G}(f)$,  ${\bf R}_{n}(f)$ and ${\bf R}_{n_{v}}^d(f)$ reduce to
${\bf g}(f)$, ${\bf R}_{\epsilon}(f)=N_0{\bf I}_{N}$ and ${\bf R}_{\epsilon_{v}}^d(f)$, respectively, and we deduce from
\eqref{eq:filter structure}, \eqref{eq:filter optimal} and \eqref{eq:filter hypothese stationnaire} that:
\begin{eqnarray}
\label{eq:wMo=wMs}
{\bf w}_{o}(f)
&=&
\{(N_0/\pi_a)+(1/T)\sum_{\ell}\|{\bf g}(f-\ell/T)\|^2\}^{-1}{\bf g}(f)\\
\label{eq:wMsc}
{\bf w}_{sc}^d(f)
&=&
\{1/\pi_a + {{\bf g}_{v}^d}^H(f)[{\bf R}_{\epsilon_{v}}^d(f)]^{-1}{\bf g}_{v}^d(f)\}^{-1}
[{\bf R}_{\epsilon_{v}}^d(f)]^{-1}{\bf g}_{v}^d(f).
\end{eqnarray}
Inserting \eqref{eq:wMo=wMs} in \eqref{eq:y optimal} and \eqref{eq:wMsc} in \eqref{eq:y structure}, it is straightforward to show that in the general case of a frequency selective SOI channel, $y_{sc}(k)$ and $y_{o}(k)$ are generally different.
This establishes the suboptimality of the sc-LMMSE receiver and, consequently, of most LMMSE receivers considered in the literature.
Nevertheless, a more detailed analysis of $y_{sc}(k)$ and $y_o(k)$ allows us to identify specific conditions under which these two outputs are equal. In particular, $y_{sc}(k)=y_o(k)$, and hence the receiver in \eqref{eq:wMsc}, is optimal, if one of the following two conditions is satisfied:
\begin{eqnarray}
\label{eq:condtion a}
&\mbox{(a)}& 
{\bf h}(f) = {\bf h}
\\
\label{eq:condtion b}
&\mbox{(b)} &
\mbox{The bandwidth of}\ v(t) \ \mbox{is included in}\ \Delta.
\end{eqnarray}
%
\section{SINR at the output of the LMMSE receivers}
\label{sec:SINR at the output of the LMMSE receivers}
%
In this section, we compute the general expression of the SINR at the output of the LMMSE receivers introduced  in Section \ref{sec:M-input LMMSE receivers}, before decision.
%
\subsection{Generic output of the $M$-input LMMSE receiver before decision}
\label{sec:Generic output of the M-input LMMSE receiver before decision}
%
We deduce from \eqref{eq:y structure}, \eqref{eq:y optimal} and \eqref{eq:y stationnaire} 
that the generic output of the $M$-input MMSE receivers considered in this paper can be written as
\begin{equation}
\label{eq:y generic}
y_{M_g}(k)
=\int  {\bf w}_{M_g}^H(f) {\bf x}_{M_g}(f)  e^{j2\pi fkT}df
\end{equation}
where it is easy to verify from \eqref{eq:model M} and \eqref{eq:model v} that
\begin{equation}
\label{eq:xM generic}
{\bf x}_{M_g}(f)
=
\sum_{\ell}a_l  e^{-j2\pi f\ell T}{\bf g}_{M_g}(f)   + {\bf n}_{M_g}(f),
\end{equation}
where
$(y_{M_g}(k),{\bf w}_{M_g}(f),{\bf x}_{M_g}(f),{\bf g}_{M_g}(f),{\bf n}_{M_g}(f))$ denotes
$(y_{sc}(k),{\bf w}_{sc}^d(f),{\bf x}_{v}(f),{\bf g}_{v}(f),{\bf n}_{v}(f))$,
$(y_{o}(k),{\bf w}_{o}(f),{\bf x}(f),{\bf g}(f),{\bf n}(f))$,
$(y_{M_s}(k),{\bf w}_{M_s}(f),{\bf x}_{M}(f),{\bf g}_{M}(f),{\bf n}_{M}(f))$
for the sc, o and s $M$-input receivers introduced in Section
\ref{sec:M-input LMMSE receivers},
respectively.
Inserting \eqref{eq:xM generic} into \eqref{eq:y generic}, we obtain:
\begin{eqnarray}
\nonumber
y_{M_g}(k)
&=&
a_k \int {\bf w}_{M_g}^H(f){\bf g}_{M_g}(f)df
+
\sum_{\ell \ne k}a_{\ell} \int {\bf w}_{M_g}^H(f){\bf g}_{M_g}(f)e^{j2\pi f (k\!-\!\ell)T}df
\\
\label{eq:y Mg}
&+&
\int {\bf w}_{M_g}^H(f){\bf n}_{M_g}(f)e^{j2\pi f kT}df
\pardef 
a_k u_{M_g}+e_{M_g}(k),
\end{eqnarray}
where it is easy to verify that $u_{M_g}(0)$ is a positive real-valued quantity and where $e_{M_g}(k)$ is the contribution of the Intersymbol Interference (ISI), the MUI and the background noise in $y_{M_g}(k)$. 
%
\subsection{Generic SINR at the output of the $M$-input LMMSE receiver before decision}
\label{sec:Generic SINR at the output of the M-input LMMSE receiver before decision}
%
The BER at the output of the generic receiver ${\bf w}_{M_g}^H(f)$ is directly linked to the SINR in $y_{M_g}(k)$,
denoted by  SINR$_{M_g}(k)$.
Using the fact that the quantities
$a_k u_{M_g}(0)$ and $e_{M_g}(k)$
are uncorrelated, we deduce that SINR$_{M_g}(k)$ can be written as
\begin{equation}
\label{eq:SINR generic k}
{\rm SINR}_{M_g}(k)
=
\frac{\pi_a u_{M_g}^2(0)}
{\e(|e_{M_g}(k)|^2)}.
\end{equation}
In the presence of MUI, the CT output
 $y_{M_g}(t)$ is SO cyclostationary,
which implies that $\e(|y_{M_g}^2(k)|)$ has Fourier series expansions given by \cite{Chevalier2018}.
\begin{equation}
\label{eq:E module y carre}
\e(|y_{M_g}^2(k)|)
=
\sum_{\alpha_i}e^{j2\pi \alpha_i kT}\int r_{y_{M_g}}^{\alpha_i}(f)df
=\sum_{\alpha_i}\int r_{y_{M_g}}^{\alpha_i}(f)df
\end{equation}
where $\alpha_i=i/T$, $i\in \mathbb{Z}$ denote the SO cyclic frequencies of $y_{M_g}(t)$,
whereas $ r_{y_{M_g}}^{\alpha_i}(f)$ is the Fourier transform of the cyclic correlation functions
$ r_{y_{M_g}}^{\alpha_i}(\tau)$ of $y_{M_g}(t)$ for the delay $\tau$ and the cyclic frequency $\alpha_i$. 
Moreover, as $y_{M_g}(t)$ is the output of the filter ${\bf w}_{M_g}^H(f)$
whose input is ${\bf x}_{M_g}(t)$, we can write
\begin{equation}
\label{eq:r gamma}
r_{y_{M_g}}^{\alpha_i}(f)
=
{\bf w}_{M_g}^H(f+\alpha_i/2){\bf R}_{x_{M_g}}^{\alpha_i}(f){\bf w}_{M_g}(f-\alpha_i/2),
\end{equation}
where ${\bf R}_{x_{M_g}}^{\alpha_i}(f)$ is the Fourier transform of the cyclic correlation matrix
${\bf R}_{x_{M_g}}^{\alpha_i}(\tau)$ of ${\bf x}_{M_g}(t)$ for the delay $\tau$ and the cyclic frequency $\alpha_i$.
\eqref{eq:E module y carre} implies that ${\rm SINR}_{M_g}(k)$ does not depends on $k$ and we get:
\begin{equation}
\label{eq:SINR generic}
{\rm SINR}_{M_g}
=
\frac{\pi_a [ \int {\bf w}_{M_g}^H(f){\bf g}_{M_g}(f)df]^2}
{\sum_{\alpha_i}\int{\bf w}_{M_g}^H(f+\alpha_i/2){\bf R}_{x_{M_g}}^{\alpha_i}(f){\bf w}_{M_g}(f-\alpha_i/2)df
-\pi_a [ \int {\bf w}_{M_g}^H(f){\bf g}_{M_g}(f)df]^2}.
\end{equation}
Furtheremore, adapting the proof given in \cite{Chevalier2024} for WL MMSE receivers, the expression \eqref{eq:SINR generic} reduces to 
\begin{equation}
\label{eq:SINR optimal}
{\rm SINR}_{1_o}
=
\frac{\pi_a}{N_0 T \int_{\Delta}{\bf e}_1^T{\bf C}_{o}^d(f){\bf e}_1df}-1.
\end{equation}
for the o-LMMSE receiver.
%
\section{SINR analysis for zero and $P$ MUI}
\label{sec:SINR analysis for zero and P MUI}
\subsection{Total noise model  and SO statistics} 
\label{sec:Total noise model  and SO statistics}
%
We assume in this section that the total noise ${\bf n}(t)$ is composed of a background noise and at most $P$ MUI, which generates the observation model 
\eqref{eq:model}.
In this context, the purpose of this section is to compute and compare the SINR at the output of the previous LMMSE receivers. From \eqref{eq:model}, for the sc and o-LMMSE and $s$ $M$-input LMMSE receivers $(M = 1, 2, 3)$,
${\bf R}_{x_{M_g}}^{\alpha_i}(f)$ appearing in \eqref{eq:SINR generic} can be written, from \cite{Chevalier2018}, as
\begin{equation}
\label{eq:R alpha}
{\bf R}_{x_{M_g}}^{\alpha_i}(f)
=
\frac{\pi_a}{T}[{\bf g}_{M_g}(f+\alpha_i/2){\bf g}_{M_g}^H(f\!-\!\alpha_i/2)
+\sum_{p=1}^P{\bf g}_{p,M_g}(f+\alpha_i/2){\bf g}_{p,M_g}^H(f-\alpha_i/2)]
+{\bf R}_{\epsilon_{M_g}}^{\alpha_i}(f),
\end{equation}
where for $p=1,..,P$, ${\bf g}_{p,M_g}(f) = {\bf g}_{p}(f)$
for $g = o$ and $M=1$ or $g = s$ and $M = 1, 2, 3$, whereas
${\bf g}_{p,M_g}(f) = {\bf g}_{p,v}(f)$ for $g = sc$ and $M = 1$.
Moreover, ${\bf R}_{\epsilon_{M_g}}^{\alpha_i}(f)$ is such that:
\begin{eqnarray}
\label{eq:Ru alpha sc 1}
{\bf R}_{\epsilon_{M_{sc}}}^{\alpha_i}(f)
&=&
N_0|v(f)|^2 \delta(\alpha_i){\bf I}_{MN};\ \  \mbox{for}\ M=1
\\
\label{eq:Ru alpha o}
{\bf R}_{\epsilon_{M_o}}^{\alpha_i}(f)
&=&
{\bf R}_{\epsilon_{M_s}}^{\alpha_i}(f)\!=\!N_0 \delta(\alpha_i){\bf I}_{MN}; 
\mbox{for}\ M=1
\\
\label{eq:R2  alpha}
{\bf R}_{{\epsilon_{M}}_s}^{\alpha_i}(f)
&=&
{N_0\delta(\alpha_i){\bf I}_{2N}
+N_0\delta(\alpha_i+1/T){\bf J}_{1}^T}
+N_0\delta(\alpha_i-1/T){\bf J}_{1} ; \mbox{for}\ M\!=\!2
\\
\nonumber
{\bf R}_{{\epsilon_{M}}_s}^{\alpha_i}(f)
&=&
N_0\delta(\alpha_i){\bf I}_{3N}
+N_0\delta(\alpha_i+2/T){\bf J}_{2}^T
+
N_0\delta(\alpha_i+1/T){\bf J}_{3}^T
+N_0\delta(\alpha_i-1/T){\bf J}_{3}
\\
\label{eq:R3  alpha}
&+&
N_0\delta(\alpha_i-2/T){\bf J}_{2} ; \ \mbox{for}\ M=3,
\end{eqnarray}
where ${\bf J}_{1}$ is the $2N \times 2N$ exchange matrix $\left( \begin{array}{cc}
\!\!{\bf 0}\!  &\! {\bf I}\!\!\\
\!\! {\bf I}\! &\! {\bf 0}\!\!\\
\end{array}\right)$ and
${\bf J}_2$ and ${\bf J}_3$
are the $3N \times3N$ matrices defined by           
\begin{equation}
\label{eq:def J}
{\bf J}_{2}
=
\left( \begin{array}{ccc}
\!\!{\bf 0}\! & {\bf 0}\! &\! {\bf 0}\!\!\\
\!\!{\bf 0}\! &\! {\bf 0}\! &\! {\bf 0}\!\!\\
\!\!{\bf 0}\! &\! {\bf I}\! &\! {\bf 0}\!\!
\end{array}\right)
\ \mbox{and}\
{\bf J}_{3}
=
\left( \begin{array}{ccc}
\!\!{\bf 0}\! &\! {\bf I} \!&\! {\bf 0}\!\!\\
\!\!{\bf 0}\! &\! {\bf 0} \!&\! {\bf 0}\!\!\\
\!\!{\bf I}\! &\! {\bf 0} \!&\! {\bf 0}\!\!
\end{array}\right).
\end{equation}
%
\subsection{SINR analysis in the absence of interference} 
\label{sec:SINR analysis in the absence of interference}
%
To further characterize and compare the performance of the sc, o and s-$M$-input LMMSE receivers ($M=1,2,3$), we consider in this section a scenario without MUI and a two-tap frequency-selective propagation channel given by
\begin{equation}
\label{eq:canal bitrajet}
{\bf h}(t)=\mu_1 \delta(t){\bf h}_1+\mu_2 \delta(t-\tau){\bf h}_2,
\end{equation}
where $\mu_1$ and $\mu_2$, with ($0\leq\mu_2\leq\mu_1$), denote the amplitudes of the first and second SOI paths, respectively, $\tau$ is the delay between the two paths, and ${\bf h}_1$ and ${\bf h}_2$ are the corresponding channel vectors, which may be either deterministic or random. Their respective components are denoted by $h_1(i)$ and $h_2(i)$, $i=1,\ldots,N$, and satisfy
${\rm E}{|h_1(i)|^2}={\rm E}{|h_2(i)|^2}=1$.

Substituting \eqref{eq:canal bitrajet} into \eqref{eq:model}, we obtain the following expression for the mean SOI symbol energy per antenna $E_s$:
\begin{equation}
\label{eq:energie bitrajet}
E_s
=
\frac{\pi_a}{N}\int \e (\|{\bf g}^2(t)\|)dt
=
\pi_a \mu_1^2
\left[1+\beta^2+2\beta{\rm Re}\left(\e\left({\bf h}_1^H{\bf h}_2/N\right)\right) r_v(\tau)\right],
\end{equation}
where $\beta\pardef\mu_2/\mu_1$ is the channel selectivity coefficient and
$r_v(t)\pardef v(t)\otimes v^*(-t)$.
Let
$\alpha_{1,2}
\pardef
{\bf h}_1^H{\bf h}_2
/(|{\bf h}_1||{\bf h}_2|)
\pardef
\alpha e^{i\phi}$
denote the spatial correlation coefficient between ${\bf h}_1$ and ${\bf h}_2$ and assume
in the random-channel case that ${\bf h}_1$ and ${\bf h}_2$ are statistically independent.
It follows that \eqref{eq:energie bitrajet} reduces to 
\begin{eqnarray}
\label{eq:energie bitrajet det}
E_s
&=&
\pi_a\mu_1^2
[1+\beta^2+2\alpha\beta \cos (\phi) r_v(\tau)],
\ \
\mbox{in the deterministic case,}	
\\		
\label{eq:energie bitrajet aleatoire}
&=&
\pi_a\mu_1^2(1+\beta^2),
\hspace{3.3cm}
\mbox{in the random case.}	
\end{eqnarray}
For a propagation channel with no delay spread ($\beta = 0$), the five receivers sc, o and s $M$-input LMMSE receivers ($M = 1, 2,3)$ are equivalent and the expected value, $\e[{\rm SINR}_{M_g}]$, of the SINR at the output of these receivers,
which corresponds to SINR$_{M_g}$ for deterministic channels, is maximal and given by
\begin{equation}
\label{eq:Esperance SINR monotrajet}
\e[{\rm SINR}_{M_g}]
= \frac{N E_s}{N_0} \pardef  \epsilon_s.
\end{equation}
Moreover, these five LMMSE receivers are also equivalent for an arbitrary propagation channel when $v(t)$ is a square-root raised-cosine filter with zero roll-off.

Except for these two cases, the sc-LMMSE receiver in \eqref{eq:filter structure} is suboptimal with respect to the o-LMMSE receiver. Computer simulations show that this suboptimality increases with the SOI bandwidth and, consequently, with the roll-off factor for raised-cosine filters. It also increases with the channel selectivity coefficient 
$\beta$ and with the magnitude $\alpha$ of the spatial correlation coefficient between ${\bf h}_1$ and ${\bf h}_2$. Furthermore, the performance loss of the sc-LMMSE receiver is more pronounced for deterministic channels than for random channels.
To quantify and illustrate the suboptimality of the sc-LMMSE receiver, we now assume that $(\alpha,\phi,\tau)$ are uniformly\footnote{The uniformity of $\alpha$ is approximately achieved by an uniform linear array 
of 3 sensors spaced half a wavelength apart, in which the elevation of the first path is $90^{°}$ and those of the second path is uniformly distributed between $48^{°}$ and $90^{°}$.} distributed over
$[0,1]\times[0,2\pi]\times [0,T]$.
Under these assumptions, with $N=3$ and $\epsilon_s=10$dB, Fig. \ref{fig:Fonction de repartition complementaire scenario pas de MUI, roll off 1 deterministe} shows, for QPSK signals, $\omega=1$, $\beta=1$, and deterministic channels, the estimated\footnote{All the CCDF and BER presented in this paper are estimated from $10^5$ independent Monte-Carlo experiments.} complementary cumulative distribution function (CCDF):
Pr[(SINR$_{M_g}/\epsilon_s){\rm dB}\ge x {\rm dB}]\pardef{P}_{M_g}(x)$ as a function of $x$ (dB).
The results clearly demonstrate the suboptimality of the sc-LMMSE receiver and the optimality of the s-$M$-input LMMSE receivers $(M=1,2,3)$ proposed in this paper under the appropriate second-order stationarity assumption for the MUI.
To complement these results, Fig. \ref{fig:Fonction de repartition complementaire scenario pas de MUI, roll off  1 Rayleigh}  shows the corresponding CCDFs under the same assumptions, but for Rayleigh-fading channels, where $h_1(1)$ and $h_2(1)$ are independent zero-mean circular complex Gaussian distributed.
The conclusions drawn from Fig. \ref{fig:Fonction de repartition complementaire scenario pas de MUI, roll off 1 deterministe} also hold for the Rayleigh-fading case shown in Fig.  \ref{fig:Fonction de repartition complementaire scenario pas de MUI, roll off  1 Rayleigh}, although the performance loss of the sc-LMMSE receiver is less pronounced.
%
\begin{figure}[H]
\begin{center}
{\psfig{file=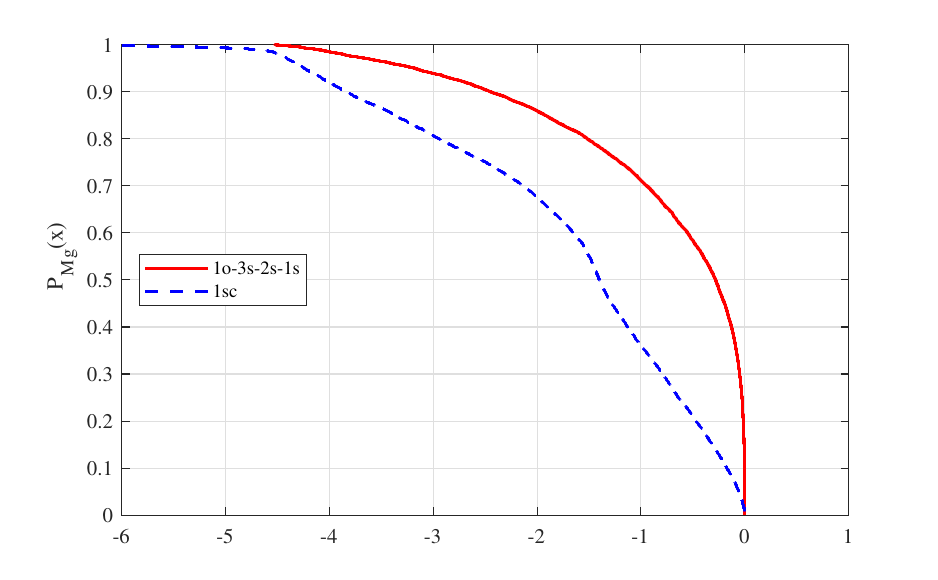,width=6in,height=3in}}
\end{center}

\vspace{-1cm}

\caption{$P_{M_g}(x)$ as a function of $x$ ($N = 3$, $\epsilon_s =$ 10 dB, $\omega = 1$, $\beta = 1$), deterministic channels, QPSK signals}
\label{fig:Fonction de repartition complementaire scenario pas de MUI, roll off 1 deterministe}

\end{figure}
%
\begin{figure}[H]
\begin{center}
{\psfig{file=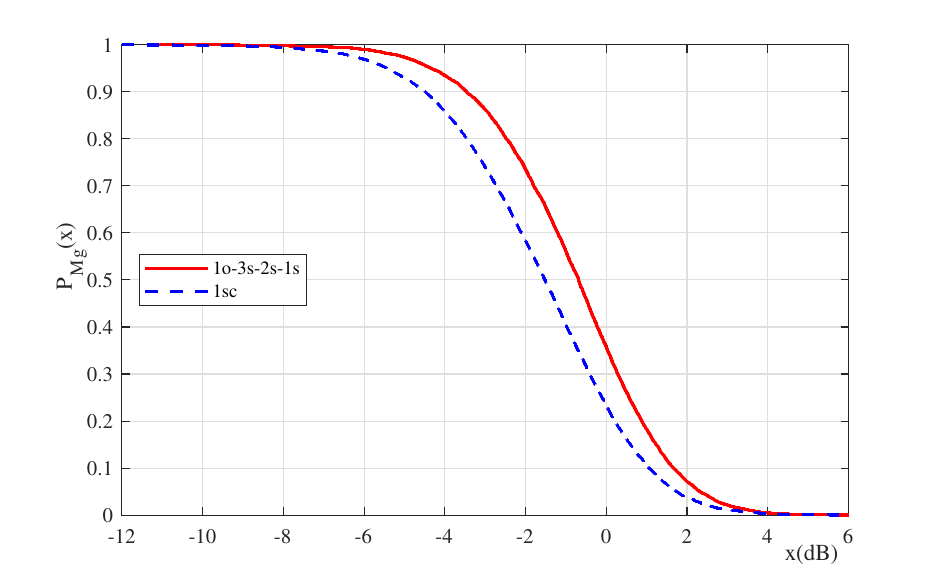,width=6in,height=3in}}
\end{center}

\vspace{-1cm}

\caption{$P_{M_g}(x)$ as a function of $x$ ($N = 3$, $\epsilon_s =$ 10 dB, $\omega = 1$, $\beta = 1$), Rayleigh channels, QPSK signals}
\label{fig:Fonction de repartition complementaire scenario pas de MUI, roll off 1 Rayleigh}

\end{figure}
\subsection{SINR analysis for $P$ MUI and channels with no delay spread} 
\label{sec:SINR analysis for P MUI and channels with no delay spread}
\subsubsection{Propagation model}
%
To analyze and compare the performance of the five LMMSE receivers introduced above in the presence of interference, we consider in this section a scenario with $P$ MUI and propagation channels without delay spread, given by
\begin{equation}
\label{eq: SOI and interference propagation channels}
{\bf h}(t)=\mu \delta(t){\bf h}
\ 
\mbox{and}
\ 
{\bf h}_p(t)=\mu_p \delta(t-\tau_p){\bf h}_p, p=1,..,P,
\end{equation}
where $\mu$ and $\mu_p$ control the amplitude of the SOI and $p$th MUI, respectively and $\tau_p$ is the delay of the $p$th MUI with respect to the SOI. 
The vectors ${\bf h}$ and ${\bf h}_p$, random or deterministic, with components $h(i)$ and $h_p(i)$, $i=1,\hdots,N$, respectively and such that 
$\e[|h(i)|^2] = \e[|h_p(i)|^2] = 1$, correspond to the channel vectors of the SOI and $p$th MUI, respectively. 
Similarly as 
\eqref{eq:energie bitrajet det} and
\eqref{eq:energie bitrajet aleatoire}, the mean SOI and MUI energy per antenna, $E_s$ and $E_p$ respectively are given by $E_s=\pi_a\mu^2$ and $E_p=\pi_{a_p}\mu_p^2$, where
$\pi_{a_p} \pardef \e|a_{p,k}^2|$. We then denote by $\epsilon_s$  and $\epsilon_p$ the quantities $\epsilon_s \pardef E_s \e(\|{\bf h}\|^2)/N_0=N E_s/N_0$ 
and $\epsilon_p \pardef E_p \e (\|{\bf h}_p\|^2)/N_0=N E_p/N_0$.
%
\subsubsection{Deterministic channels and SRRC filters with zero roll-off}
\label{sec:Deterministic channels and SRRC filters with zero roll-off}
Under the assumptions above, analytically tractable expressions for SINR$_{M_g}$, defined in \eqref{eq:SINR generic}, can be obtained only when $v(t)$ is a square-root raised-cosine (SRRC) filter with zero roll-off, i.e., $\omega=0$, which is assumed throughout this subsection. For nonzero roll-off, SINR$_{M_g}$ can only be evaluated numerically through computer simulations; these results are presented in Subsection \ref{sec:Deterministic channels and SRRC filters with arbitrary roll-off}.
Moreover, we assume in this subsection deterministic channels and we denote by 
$\alpha_{s,p}\pardef {\bf h}^H{\bf h}_p/(\|{\bf h}\|\|{\bf h}_p\|)= |\alpha_{s,p}| e^{j \phi_{s,p}}$ the spatial correlation coefficient between the SOI and the $p$th MUI, such that $0\le |\alpha_{s,p}| \le 1$,

For $P$ MUI with $1 \le P<N$, we obtain the following expressions from \eqref{eq:SINR generic} after 
cumbersome derivations, whose proof of 
\eqref{eq:SINRo sc  s zero roll off P>1} is given in the Appendix:
\begin{eqnarray}
\label{eq:SNRo sc  s zero roll off}
{\rm SINR}_{1_{sc}}
&=&
{\rm SINR}_{1_{o}}
=
{\rm SINR}_{1_{s}}
=
{\rm SINR}_{2_{s}}
=
{\rm SINR}_{3_{s}}
\\
\label{eq:SINRo sc  s zero roll off P=1}
&=&
\epsilon_s\left(
1-\frac{|\alpha_{s,1}|^2\epsilon_1}{1+\epsilon_1}
\right),\ \  \mbox{for}\ P=1
\\
\label{eq:SINRo sc  s zero roll off P>1}
&\approx&
\epsilon_s(1-|\alpha_{s,P}|^2),\hspace{1.1cm}  \mbox{for}\ P\ \mbox{strong MUI, i.e.,}\ \epsilon_p  \gg 1, p=1,...,P,
\end{eqnarray}
with 
$|\alpha_{s,P}|=\|{\bf \Pi}_P{\bf h}\|/\|{\bf h}\|\in [0,1]$
where
${\bf \Pi}_P\pardef  {\bf H}_P({\bf H}_P^H{\bf H}_P)^{-1}{\bf H}_P^H$ and ${\bf H}_P\pardef [{\bf h}_1,..., {\bf h}_p,..,{\bf h}_P]$
is the spatial correlation between the SOI and the plane generated by the $P$ MUI.
This result extends the result proven in \cite{Lin1982} for the case of spatial filtering.
%
\subsubsection{Deterministic channels and flat pulse shape $v(f)$ of bandwidth $2/T$}
\label{sec:Deterministic channels and v(f) bandwidth 2/T}
%
To derive interpretable analytical expressions for the SINRs that allow us to demonstrate the suboptimality of the conventional sc-LMMSE receiver and the near-optimality of the $M$-input LMMSE receivers $(M=2,3)$ in the presence of signals whose bandwidth exceeds $\Delta$, 
we consider the following idealized pulse-shaping filter: $v(f)=\sqrt{T/2},\mathds{1}_{[-1/T,1/T]}(f)$.
In Subsection \ref{sec:Deterministic channels and SRRC filters with arbitrary roll-off}, we show that these analytical expressions also provide useful insights into the SINRs obtained with SRRC filters having a roll-off factor of one.
After lengthy but straightforward algebraic derivations based on \eqref{eq: SINR by output power}-\eqref{eq:calcul de puissance bruit blanc},
 the following expressions for the SINRs of the sc- and o-LMMSE receivers in the single-MUI case $(P=1)$ are derived:
%
%
\begin{eqnarray}
\label{eq:SINRo sc 2/T P=1}
{\rm SINR}_{1_{sc}}
&=&
\epsilon_s
\left(
1-|\alpha_{s,1}|^2\frac{\epsilon_1 \cos^2(\frac{\pi\tau_1}{T})}{1+\epsilon_1 \cos^2(\frac{\pi\tau_1}{T})}
\right)
\approx
\epsilon_s
\left(
1-|\alpha_{s,1}|^2
\right),\
\mbox{for}\ \epsilon_1  \gg 1,
\\
\label{eq:SINRo o 2/T P=1}
{\rm SINR}_{1_{o}}
&=&
\epsilon_s
\left(
1-|\alpha_{s,1}|^2\frac{\epsilon_1 \cos^2(\frac{\pi\tau_1}{T})}{1+\epsilon_1}
\right)
\approx
\epsilon_s
\left(
1-|\alpha_{s,1}|^2 \cos^2\frac{\pi\tau_1}{T}
\right),\
\mbox{for}\ \epsilon_1  \gg 1.
\end{eqnarray}
Clearly from \eqref{eq:SINRo sc 2/T P=1} and \eqref{eq:SINRo o 2/T P=1}, we see that ${\rm SINR}_{1_{sc}} <{\rm SINR}_{1_{o}}$ except
for synchronized SOI and MUI, (i.e., for $\tau_1=kT$), for which
${\rm SINR}_{1_{o}}
= {\rm SINR}_{1_{sc}}
=\epsilon_s \left(1-|\alpha_{s,1}|^2\frac{\epsilon_1}{1+\epsilon_1}\right)$.

In the multi-MUI case ($P>1$), the derivation of ${\rm SINR}_{1_{sc}}$ and ${\rm SINR}_{1_{o}}$ are much more involved
and we were only able to prove that for orthogonal vectors ${\bf h}_1,..., {\bf h}_p,..,{\bf h}_P$ we have:
\begin{eqnarray}
\label{eq:SINR sc h orthogonal}
{\rm SINR}_{1_{sc}}
&=&
\epsilon_s
\left(
1-\sum_{p=1}^P\frac{|\alpha_{s,p}|^2 \epsilon_p\cos^2(\frac{\pi \tau_p}{T})}{1+ \epsilon_p\cos^2(\frac{\pi \tau_p}{T})}
\right),\\
\label{eq:SINR o h orthogonal}
{\rm SINR}_{1_{o}}
&=&
\epsilon_s
\left(
1-\sum_{p=1}^P\frac{|\alpha_{s,p}|^2 \epsilon_p\cos^2(\frac{\pi \tau_p}{T})}{1+ \epsilon_p}
\right),
\end{eqnarray}
which confirms once again that ${\rm SINR}_{1_{sc}} \le {\rm SINR}_{1_{o}}$.
Furthermore, we verify that for any vector ${\bf h}_1,..., {\bf h}_p,..,{\bf h}_P$ and for SOI and MUI in phase opposition, i.e., $\tau_p = T/2+k_pT$, $p=1,..,P$, we obtain
${\rm SINR}_{1_{o}}
= {\rm SINR}_{1_{sc}}
=\epsilon_s$, meaning the MUI is completely eliminated.

For $M$input s-LMMSE-receivers $(M=1,2,3)$, only the $P=1$ case can yield interpretable closed-form expressions of the SINR.
In particular, we obtain:
\begin{eqnarray}
\label{eq:SINR s a}
{\rm SINR}_{1_{s}}
&=&
{\rm SINR}_{2_{s}}
=
{\rm SINR}_{3_{s}}
=
{\rm SINR}_{sc}={\rm SINR}_{o}=
\epsilon_s
\frac{1}{1+\epsilon_1},\ \
 \mbox{for}\ |\alpha_{s,1}|=1, \tau_1=kT,
 \\
 \label{eq:SINR s b}
 {\rm SINR}_{1_{s}}
&=&
\epsilon_s
\frac{1}{1+\epsilon_1 \cos^2\frac{\pi\tau_1}{T}},\hspace{5.7cm} \mbox{for}\ |\alpha_{s,1}|=1
\end{eqnarray}
In the specific case $|\alpha_{s,1}|=1$, $\epsilon_1 \gg 1$, we get:
\begin{eqnarray}
\label{eq:SINR sc c}
 {\rm SINR}_{1_{sc}}
 &\approx&
 \epsilon_s 
 \frac{1}{\epsilon_1\cos^2\frac{\pi\tau_1}{T}}, \ \mbox{for}\  \tau_1 \ne T/2+kT,
 \\
\label{eq:SINR o c}
  {\rm SINR}_{1_{o}}
 &\approx&
 \epsilon_s 
\left(\sin^2\frac{\pi\tau_1}{T}+ \frac{\cos^2\frac{\pi\tau_1}{T}}{\epsilon_1}\right),
\\
\label{eq:SINR 1s c}
{\rm SINR}_{1_{s}}
&\approx&
 \epsilon_s 
 \frac{1}{\epsilon_1\cos^2\frac{\pi\tau_1}{T}}, \ \mbox{for}\  \tau_1 \ne T/2+kT,
\\
\label{eq:SINR 2s c}
{\rm SINR}_{2_{s}}
&\approx&
\epsilon_s\left( \sin^2\frac{\pi\tau_1}{T}
-\frac{3}{2\epsilon_1}\cos^2\frac{\pi\tau_1}{T}\right), \ \mbox{for}\  \tau_1 \ne kT,
\\
\label{eq:SINR 3s c}
{\rm SINR}_{3_{s}}
&\approx&
\epsilon_s\left( \sin^2\frac{\pi\tau_1}{T}
+
\frac{1}{\epsilon_1^2}\frac{(1+\cos\frac{2\pi\tau_1}{T})(5-\cos\frac{2\pi\tau_1}{T})}{(1-\cos \frac{2\pi\tau_1}{T})}\right), \ \mbox{for}\  \tau_1 \ne kT,
\end{eqnarray}
where you can cheek that ${\rm SINR}_{sc}\approx {\rm SINR}_{1_s}\le  {\rm SINR}_{2_s}\le  {\rm SINR}_{3_s}< {\rm SINR}_{o}$. 
Furthermore, we get for $(|\alpha_{s,1}|,\tau_1) \ne (1,kT)$:
\begin{eqnarray}
\label{eq:SINR 1s}
{\rm SINR}_{1_{s}}
&\approx&
\epsilon_s
(1-|\alpha_{s,1}|^2)
\\
\label{eq:SINR 2s}
{\rm SINR}_{2_{s}}
&\approx&
\epsilon_s
\frac{\left(
1-|\alpha_{s,1}|^2\left(\frac{3+ \cos(\frac{2\pi\tau_1}{T})}{4}
\!\right)\right)^2}
{\left(
1-|\alpha_{s,1}|^2\left(\frac{7+ \cos(\frac{2\pi\tau_1}{T})}{8}
\!\right)\right)}
\\
\label{eq:SINR 3s}
{\rm SINR}_{3_{s}}
&\approx&
\epsilon_s
\frac{\left(
1-|\alpha_{s,1}|^2\left(\frac{2+ \cos(\frac{2\pi\tau_1}{T})}{3}
\!\right)\right)^2}
{\left(
1-|\alpha_{s,1}|^2\left(\frac{7+ 2\cos(\frac{2\pi\tau_1}{T})}{9}
\!\right)\right)},
\end{eqnarray}
showing again that  ${\rm SINR}_{sc}\approx {\rm SINR}_{1_s}\le  {\rm SINR}_{2_s}\le  {\rm SINR}_{3_s}< {\rm SINR}_{o}$.
Finally, note that we have proven theoretically that all the SINR do not depend on the phases $\phi_{s,p}$.
%
\subsubsection{Deterministic channels and SRRC filters with arbitrary roll-off}
\label{sec:Deterministic channels and SRRC filters with arbitrary roll-off}
%
To compare the SINRs of the sc-, o-, and s-$M$-input LMMSE receivers $(M=1,2,3)$ for arbitrary roll-off $\omega$, we adopt a statistical approach. We continue to consider deterministic channels and a single MUI signal $(P=1)$, and assume that $(|\alpha_{s,1}|,\phi_{s,1},\tau_1)$ is  uniformly\footnote{The uniformity of $|\alpha_{s,1}|$ is approximately achieved by an uniform linear array 
of 3 sensors spaced half a wavelength apart, in which the elevation of the SOI is $90^{°}$ and those of the MUI are uniformly distributed between $48^{°}$ and $90^{°}$.} randomly distributed on 
$ [0,1] \times [0, 2\pi]\times [0, T]$ and we choose $\epsilon_s = 10$ dB and $\epsilon_1 = 20$ dB.
Under these assumptions, Fig. \ref{fig:Fonction de repartition complementaire scenario 1 MUI, roll off 1 deterministe} shows for $N = 3$, $\omega=1$, 
the variations of Pr[(SINR$_{M_g}/\epsilon_s){\rm dB}\ge x {\rm dB}]\pardef{P}_{M_g}(x)$ as a function of $x$ (dB) for the sc, o and s $M$-input LMMSE receivers ($M = 1, 2,3$).
Fig. \ref{fig:Fonction de repartition complementaire scenario 2 MUI, roll off 1 deterministe} shows the same variations but with $P=2$ equipowered MUI 
($\epsilon_1 =\epsilon_2= 20$ dB) where $(|\alpha_{s,1}|,|\alpha_{s,2}|,\phi_{s,1},\phi_{s,2},\tau_1,\tau_2)$ are assumed randomly uniformly distributed on
$ [0,1]^2 \times [0, 2\pi]^2 \times [0, T]^2$. 
These two figures show that the SINR given by the sc and s-one input LMMSE receivers are very close, similarly as in our theoretical analysis derived for $\omega=0$. We note the largely sub-optimality of these two LMMSE receivers, a more pronounced sub-optimality for two MUI than for a single MUI.
These figures also show that the s-2 and 3 input LMMSE receivers are quasi-optimal.
%
\begin{figure}[H]
\begin{center}
{\psfig{file=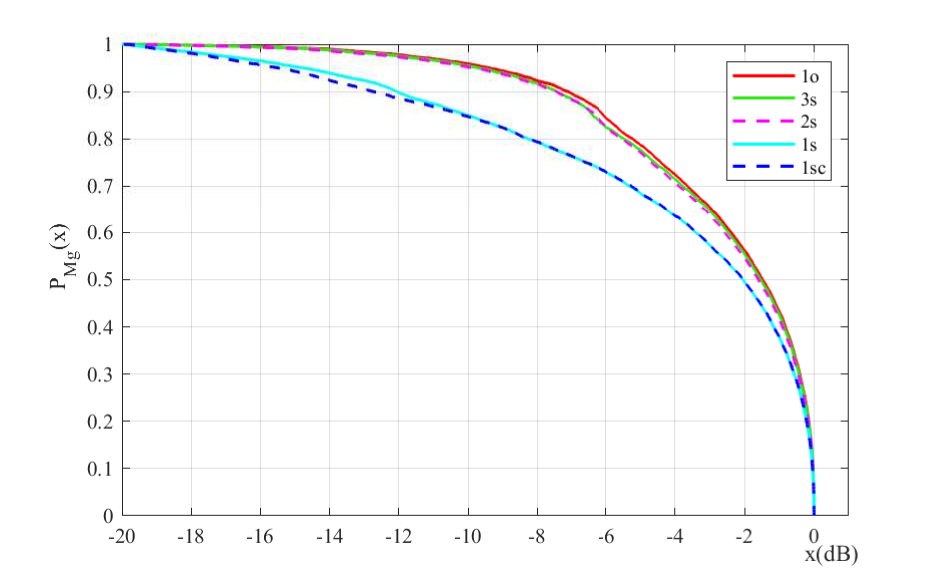,width=6in,height=3in}}
\end{center}

\vspace{-1cm}

\caption{$P_{M_g}(x)$ as a function of $x$ ($N = 3$, $P=1$, $\epsilon_s =$ 10 dB, $\epsilon_1 =$ 20 dB, $\omega = 1$), deterministic channels, QPSK signals}
\label{fig:Fonction de repartition complementaire scenario 1 MUI, roll off 1 deterministe}

\end{figure}
\begin{figure}[H]
\begin{center}
{\psfig{file=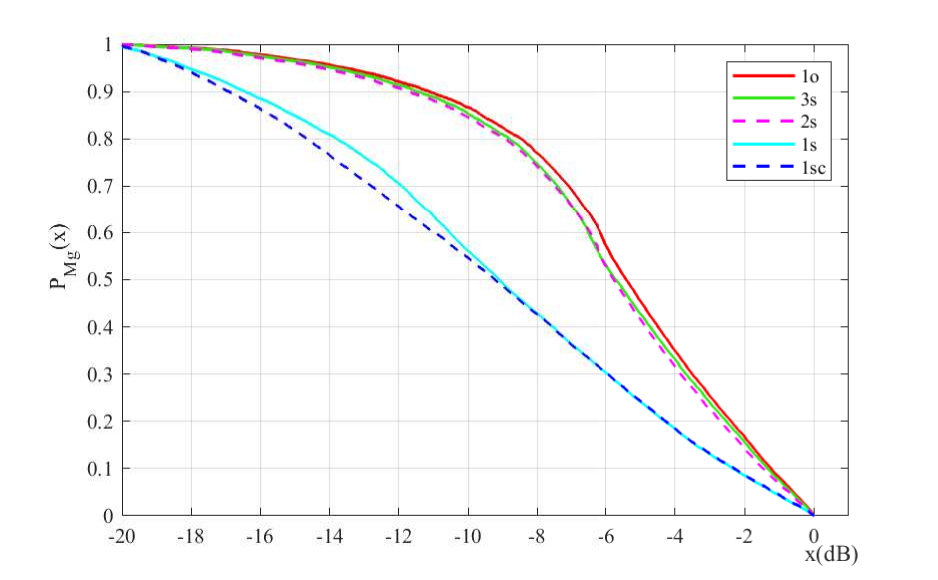,width=6in,height=3in}}
\end{center}

\vspace{-1cm}

\caption{$P_{M_g}(x)$ as a function of $x$ ($N = 3$, $P=2$, $\epsilon_s =$ 10 dB, $\epsilon_1 =\epsilon_2=$ 20 dB, $\omega = 1$), deterministic channels, QPSK signals}
\label{fig:Fonction de repartition complementaire scenario 2 MUI, roll off 1 deterministe}
\end{figure}

\subsubsection{Rayleigh channels and SRRC filters with arbitrary roll-off}
\label{sec:Rayleigh channels and SRRC filters with arbitrary roll-off}
%
To complete the previous results,
Figs. 
\ref{fig:Fonction de repartition complementaire scenario 1 MUI, roll off 1 Rayleigh}
and 
\ref{fig:Fonction de repartition complementaire scenario 2 MUI, roll off 1 Rayleigh}
show the same variations as Figs. \ref{fig:Fonction de repartition complementaire scenario 1 MUI, roll off 1 deterministe}
and 
\ref{fig:Fonction de repartition complementaire scenario 2 MUI, roll off 1 deterministe},
but
for Rayleigh fading channels,
for which $\tau_i,h(j),h_i(j)$ with $i=1,2$ and $j=1,2,3$ are independent, where $(\tau_1,\tau_2)$ is uniformly distributed in $[0,T]^2$
and $h(j),h_i(j)$ zero-mean circular Gaussian distributed.
The conclusions of Figs.
\ref{fig:Fonction de repartition complementaire scenario 1 MUI, roll off 1 deterministe}
and \ref{fig:Fonction de repartition complementaire scenario 2 MUI, roll off 1 deterministe}
hold for
Figs.  \ref{fig:Fonction de repartition complementaire scenario 1 MUI, roll off 1 Rayleigh} and
  \ref{fig:Fonction de repartition complementaire scenario 2 MUI, roll off 1 Rayleigh}.
%
\begin{figure}[H]
\begin{center}
{\psfig{file=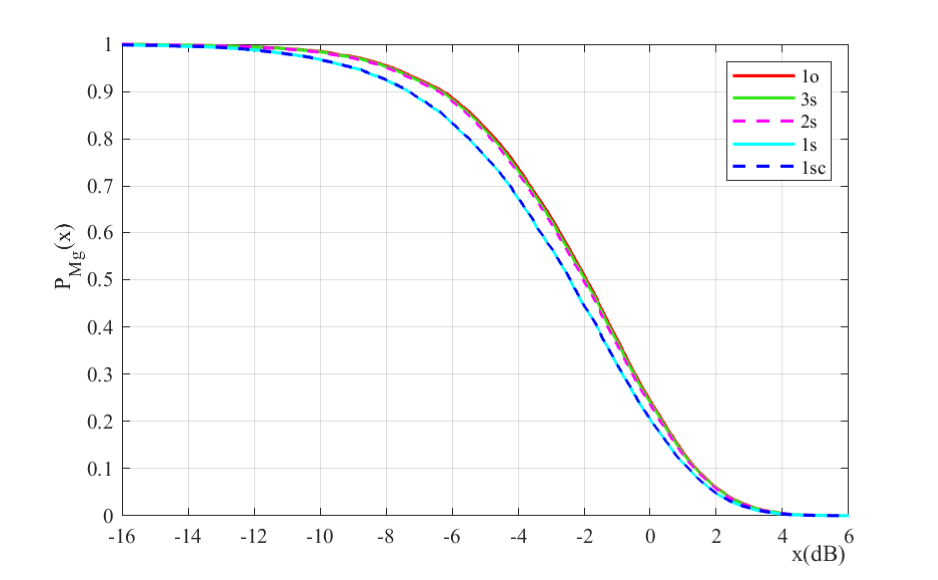,width=6in,height=3in}}
\end{center}

\vspace{-1cm}

\caption{$P_{M_g}(x)$ as a function of $x$ ($N = 3$, $P=1$, $\epsilon_s =$ 10 dB, $\epsilon_1 =$ 20 dB, $\omega = 1$), Rayleigh channels, QPSK signals}
\label{fig:Fonction de repartition complementaire scenario 1 MUI, roll off 1 Rayleigh}
\end{figure}
\begin{figure}[H]
\begin{center}
{\psfig{file=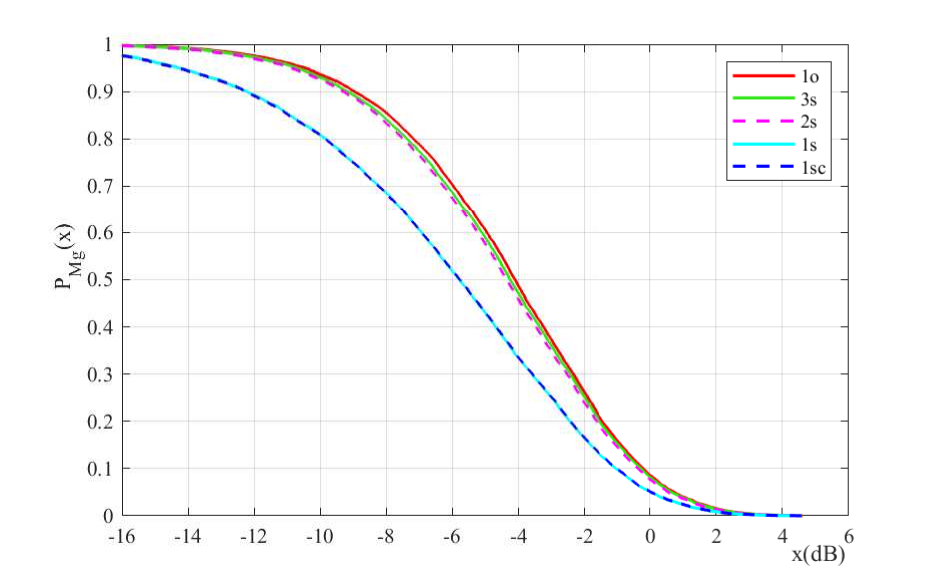,width=6in,height=3in}}
\end{center}

\vspace{-1cm}

\caption{$P_{M_g}(x)$ as a function of $x$ ($N = 3$, $P=2$, $\epsilon_s =$ 10 dB, $\epsilon_1 =\epsilon_2=$ 20 dB, $\omega = 1$), Rayleigh channels, QPSK signals}
\label{fig:Fonction de repartition complementaire scenario 2 MUI, roll off 1 Rayleigh}
\end{figure}
\section{BER at the output of the LMMSE receivers for one MUI}
\label{sec:BER at the output of the LMMSE receivers for one MUI}
In this section, we show that the main conclusions drawn in the previous section from the output SINR analysis remain valid when assessed in terms of the output BER.

\subsection{Theoretical closed-form expressions}
\label{sec:Theoretical closed-form expressions}
%
To compare the previous five LMMSE receivers from an output BER analysis, we still consider the observation model \eqref{eq:model M} which gives from
the generic output $y_{M_g}(k)$ of the different LMMSE receiver for $M_g=1_{sc},1_o,1_{s},2_{s},3_{s}$.
\begin{equation}
\label{eq:z generic bis}
y_{M_g}(k)
=
\underbrace{a_k u_{M_{g}}(0)}_{\rm SOI}
+\underbrace{\sum_{\ell \ne k}a_{\ell}u_{M_{g}}(k-\ell)}_{\rm ISI}
+\sum_{p=1}^{P}
\underbrace{\sum_{\ell }a_{p,\ell}u_{M_{g,p}}(k-\ell)}_{p{\rm th MUI}}
+\underbrace{\epsilon_{M_g}(k)}_{\rm BN}
\end{equation}
where
$u_{M_{g}}(k)\pardef \int {\bf w}_{M_g}^H(f){\bf g}_{M_g}(f)e^{j2\pi f kT}df$,
$u_{M_{g,p}}(k)\pardef  \int {\bf w}_{M_g}^H(f){\bf g}_{p,M_g}(f)e^{j2\pi f kT}df$ and
$\epsilon_{M_g}(k)\pardef  \int {\bf w}_{M_g}^H(f){\boldsymbol \epsilon}_{M}(f)e^{j2\pi f kT}df$ is the background noise (BN) component which is zero-mean circular Gaussian distributed with variance
$\sigma^2_{\epsilon_{M_{g}}}=N_0\int \| {\bf w}_{1_g}(f)\|^2df$ for the o and s one input LMMSE receivers,
$\sigma^2_{\epsilon_{M_{sc}}}=N_0\int  |v(f)|^2\|{\bf w}_{1_{sc}}(f)\|^2df$ and
$\sigma^2_{\epsilon_{M_{s}}}=\sum_{\alpha_i}\int  {\bf w}_{M_s}^H(f+\frac{\alpha_i}{2}){\bf R}_{{\epsilon_{M}}_s}^{\alpha_i}(f) {\bf w}_{M_s}(f-\frac{\alpha_i}{2})df$ for the s $M$ input LMMSE ($2,3$) receivers.

When the number of ISI and MUI terms at the output $y_{M_g}(k)$ of the $M$-input LMMSE receivers is large and when there are no dominant term in the ISI and MUI, an approximation of the central limit theorem (Lyapounov theorem \cite[th. 27.3]{Billingsley1986}) for independent non identically distributed random variables can be applied and the BER is directly deduced from the SINR. For example, for QPSK symbols $a_k$ and $a_{p,k}, p=1,..P$, we get the approximation of the BER for Gray-coded QPSK symbols derived from the optimal detector $\hat{a}_k =\arg \min_{a\in \{-1,+1,-j,+j\}}|y_{M_g}(k)-a|^2$.
\begin{equation}
\label{eq:BER CLT}
{\rm BER}_{M_g}
\approx
Q(\sqrt{{\rm SINR}_{M_g}}),
\end{equation}
with $Q(x) \pardef \int_x^{+\infty}\frac{1}{\sqrt{2\pi}}e^{-\frac{u^2}{2}}du$.
This relation \eqref{eq:BER CLT} confirms that the performance in term of output SINR and BER are equivalent.
When this number of ISI and MUI  terms is weak and/or there is a dominant term in the ISI or MUI, the approximation is no longer valid, 
we must then resort to Monte-Carlo experiments.
%
\subsection{Monte-Carlo experiments}
\label{sec:Monte-Carlo experiments}
We consider here the case of one MUI where the symbols $a_k$ and $a_{1,k}$ are QPSK $(-1,+1,-j,+j)$.
\subsubsection{One tap deterministic channels}
We consider the transmission of 1000 frames of 200 QPSK symbols and we assume, in this subsection, one tap deterministic channels which are constant over a frame and random from a frame to another. For each frame, we assume that $((|\alpha_{s,1}|,\phi_{s,1},\tau_1)$ 
are independent uniformly distributed on $[0,1]\times[0, 2\pi]\times [0,T]$. 
Under these assumptions, Fig. \ref{fig:BER as a function of epsilon deterministic one tap channel, R and QPSK signals} shows the variations of the BER given by the optimal detector at the output of the different LMMSE receivers, as a function of $\epsilon_s$ for $N=3$, $\epsilon_1/\epsilon_s=10$ dB and $\omega=1$.
 
The results of Fig. \ref{fig:BER as a function of epsilon deterministic one tap channel, R and QPSK signals} confirm, from a BER perspective, the sub-optimality of the sc and s one-input receivers and the quasi-optimality of the s 2 and 3-input LMMSE receivers. 
 %
\begin{figure}[H]
\begin{center}
{\psfig{file=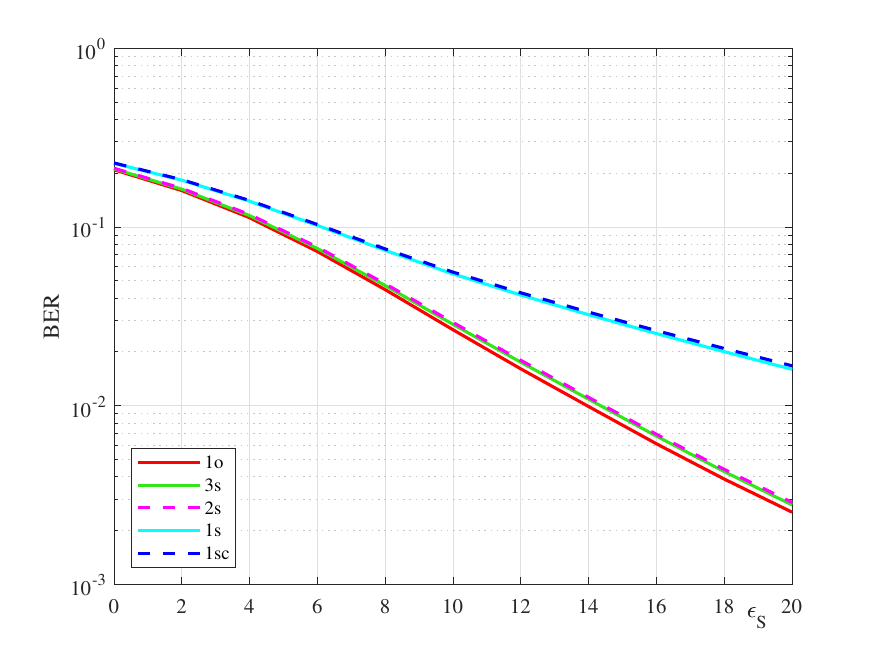,width=6in,height=3in}}
\end{center}

\vspace{-1cm}

\caption{BER as a function of $\epsilon_s$ ($N=3$, $\epsilon_1/\epsilon_s=10$ dB, $\omega=1$), deterministic one tap channel, QPSK signals.}
\label{fig:BER as a function of epsilon deterministic one tap channel, R and QPSK signals}
\end{figure}
\subsubsection{One tap Rayleigh channels}
\begin{figure}[H]
\begin{center}
{\psfig{file=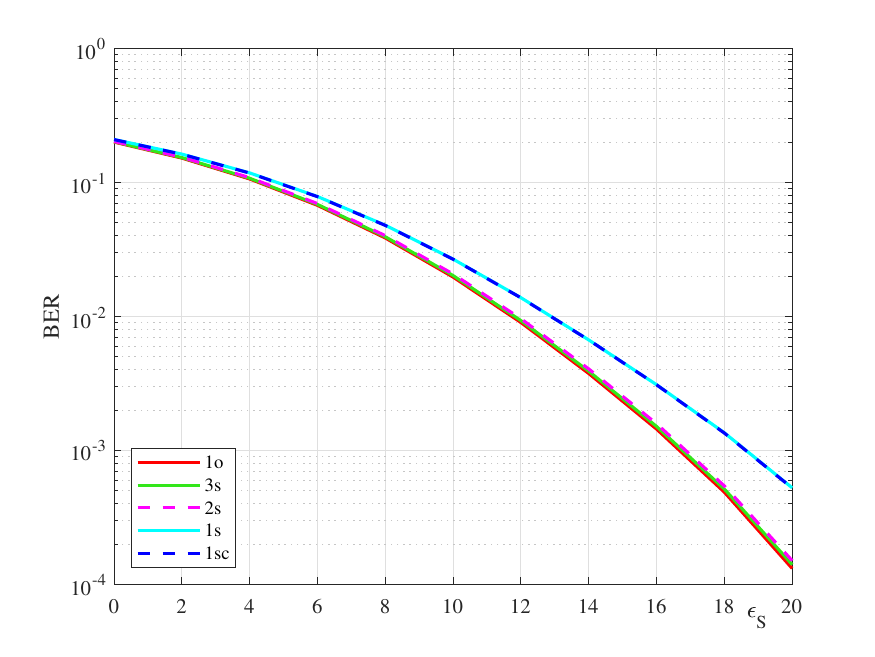,width=6in,height=3in}}
\end{center}

\vspace{-1cm}

\caption{BER as a function of $\e(\epsilon_s)$ ($N=3$, $\e(\epsilon_1)/\e(\epsilon_s)=10$ dB, $\omega=1$), Rayleigh fading one tap channel, QPSK signals.}
\label{fig:BER as a function of epsilon Rayleigh fading one tap chanel QPSK signals}
\end{figure}

To complete the previous results and under the assumptions of Fig. \ref{fig:BER as a function of epsilon deterministic one tap channel, R and QPSK signals}, 
Fig. \ref{fig:BER as a function of epsilon Rayleigh fading one tap chanel QPSK signals}
 shows the same variations as 
Fig. \ref{fig:BER as a function of epsilon deterministic one tap channel, R and QPSK signals}, but as a function of $\e(\epsilon_s)$ for Rayleigh fading channels for which $h(j)$ and $h_1(j)$, $j=1,2,3$ are zero-mean circular Gaussian independently distributed, such that $\e[\epsilon_1]/\e[\epsilon_s] = 10$ dB. The conclusions of Fig. \ref{fig:BER as a function of epsilon deterministic one tap channel, R and QPSK signals} hold for Fig. \ref{fig:BER as a function of epsilon Rayleigh fading one tap chanel QPSK signals}.
%
\section{Conclusion}
\label{sec:Conclusion}
%
This paper has provided new insights into LMMSE receivers, which are of fundamental importance in numerous current and emerging applications, both in the absence and, more importantly, in the presence of MUI. A CT framework was considered for propagation channels with and without delay spread.
We first recalled that most LMMSE receivers in the literature (denoted sc-LMMSE), are implemented at the symbol rate following a matched-filtering operation with the pulse-shaping filter. This implementation imposes a specific structural constraint but does not require knowledge of the MUI channels, contrary to what is often implicitly assumed. However, we showed that such receivers are generally suboptimal in frequency-selective channels and in the presence of MUI.
We then derived the optimal LMMSE (denoted o-LMMSE) receiver. Unlike the sc-LMMSE receiver, the o-LMMSE receiver is not subject to any structural constraint and can exploit the MUI waveforms and their cyclostationarity. However, its implementation requires knowledge or estimation of the MUI channels, which may be cumbersome in practice.
The main challenge addressed in this paper was therefore to design LMMSE receivers capable of achieving the performance of the o-LMMSE receiver without requiring knowledge of the MUI channels. To this end, two novel LMMSE receivers, referred to as the two-input and three-input LMMSE receivers, were proposed under the assumption that the MUI is SO stationary. These receivers require only knowledge or estimation of the SOI channel, which makes them particularly attractive for practical applications.
Their performance, measured in terms of the output SINR for the current symbol prior to decision, was analyzed analytically and compared with that of the sc-LMMSE and o-LMMSE receivers. For specific channel models, interpretable closed-form expressions for the output SINR were obtained. These analytical results were complemented by computer simulations for more general channel models and by a BER analysis.
They show the quasi-optimality of  the two-input and three-input LMMSE receivers.
This result provides new perspectives for the practical implementation and optimization of LMMSE receivers in the presence of MUI when only knowledge or estimation of the SOI channel is available.
%
\appendix
To derive closed-form expressions for the SINR, it is more convenient to calculate the powers $P_{\rm SOI}$, $P_{\rm ISI}$, $P_{\rm MUI}$, and $P_{\rm BN}$ of the different terms in \eqref{eq:z generic bis}, rather than directly using the compact expressions \eqref{eq:SINR generic} and \eqref{eq:SINR optimal}:

\begin{equation}
\label{eq: SINR by output power}
{\rm SINR}_{M_g}
=
\frac{P_{\rm SOI}}{P_{\rm ISI}+P_{\rm MUI}+P_{\rm BN}},
\end{equation}
where the different powers are given by
\begin{eqnarray}
\label{eq:calcul de puissance SOI}
P_{\rm SOI}
&=&
\pi_a\left(\int{\bf w}^H_{M_g}(f){\bf g}_{M_g}(f)df\right)^2,
\\
\label{eq:calcul de puissance ISI}
P_{\rm ISI}
&=&
\frac{\pi_a}{T}
\int_{\Delta}|\sum_k{\bf w}^H_{M_g}(f-\frac{k}{T}){\bf g}_{M_g}(f-\frac{k}{T})|^2df
-
\pi_a\left(\int{\bf w}^H_{M_g}(f){\bf g}_{M_g}(f)df\right)^2,
\\
\label{eq:calcul de puissance MUI}
P_{\rm MUI}
&=&
\frac{\pi_a}{T}
\sum_{p=1}^P
\int_{\Delta}|\sum_k{\bf w}^H_{M_g}(f-\frac{k}{T}){\bf g}_{p,M_g}(f-\frac{k}{T})|^2df,
\\
\label{eq:calcul de puissance bruit blanc}
P_{\rm BN}
&=&
\sum_{\alpha_i}
\int {\bf w}^H_{M_g}(f+\frac{\alpha_i}{2}){\bf R}^{\alpha}_{\epsilon_{M_g}}(f)
{\bf w}_{M_g}(f-\frac{\alpha_i}{2})df.
\end{eqnarray}
%
%
Using \eqref{eq:SNRo sc  s zero roll off}, let us calculate SINR$_{1_s}$ \eqref{eq:SINRo sc  s zero roll off P>1}, where from \eqref{eq:filter hypothese stationnaire} 
\begin{equation}
\label{eq:ws1}
{\bf w}_{1,s}(f)={c_{1_s}^d}(f)[{\bf R}_{n_1}^{0}(f)]^{-1}{\bf g}(f), 
\end{equation}
with
\begin{equation}
\label{eq:Rn0}
{\bf R}_{n_1}^{0}(f)
=
{\bf H}_P\diag(\pi_a\mu^2_1,\hdots,\pi_a\mu^2_P){\bf H}_P^H \mathds{1}_{\Delta}(f)+N_0{\bf I}_N,
\end{equation}
which, by the matrix inversion lemma, gives :
\begin{eqnarray}
\nonumber
[{\bf R}_{n_1}^{0}(f)]^{-1}
&=&
\frac{1}{N_0}
\left(
{\bf I}_N-{\bf H}_P\left[{\bf H}^H_P{\bf H}_P+\diag\left(\frac{N_0}{\pi_a\mu^2_1},\hdots,\frac{N_0}{\pi_a\mu^2_1}\right)\right]^{-1}{\bf H}^H_P
\right)
\mathds{1}_{\Delta}(f)+\frac{1}{N_0}{\bf I}_N,
\\
\label{eq:inverseRn0}
&\approx&
\frac{1}{N_0}
\left(
{\bf I}_N-{\bf H}_P({\bf H}^H_P{\bf H}_P)^{-1}{\bf H}^H_P
\right)\mathds{1}_{\Delta}(f)+\frac{1}{N_0}{\bf I}_N
=\frac{1}{N_0}{\bf \Pi}_P^{\bot}
\mathds{1}_{\Delta}(f)+\frac{1}{N_0}{\bf I}_N,
\end{eqnarray}
for $\epsilon_p=\frac{\pi_a\mu_p^2\|{\bf h}_p\|^2}{N_0} \gg 1, p=1,\hdots,P$.
Substituting the approximate expression in \eqref{eq:inverseRn0} into \eqref{eq:ws1} and carrying out the resulting calculations yields

\begin{equation}
\label{eq:ws1closedform}
{\bf w}_{1,s}(f)
\approx c {\bf \Pi}_P^{\bot}{\bf h}
\mathds{1}_{\Delta}(f)
\end{equation}
with $c\pardef N_0^{-1}\pi_a\mu[1+\epsilon(1-|\alpha_{s,P}|)^2]^{-1}\sqrt{T}$. Substituting this approximate expression for 
${\bf w}_{1,s}(f)$ into the component-power expressions \eqref{eq:calcul de puissance SOI}-\eqref{eq:calcul de puissance bruit blanc}  gives:
\begin{equation}
\label{eq:power}
P_{\rm SOI}\approx \pi_ac^2\epsilon_s^2(1-|\alpha_{s,P}|^2)^2,\
P_{\rm ISI}=0,\ P_{\rm MUI}\approx 0\ 
\mbox{and}\ 
P_{\rm BN}\approx \pi_ac^2\epsilon_s(1-|\alpha_{s,P}|^2),
\end{equation}
which proves \eqref{eq:SINRo sc  s zero roll off P>1}.
\hfill
\QED


\end{document}